# Comment on "Spin-Momentum-Locked Edge Mode for Topological Vortex Lasing, Phys. Rev. Lett. vol. 125, 013903 (2020)"

Xiao-Chen Sun[1], Xing-Xiang Wang[1], Tomohiro Amemiya[2] and Xiao Hu[1]

1) International Center for Materials Nanoarchitectonics, National Institute for Materials Science

Tsukuba 305-0044, Japan

2) Institute of Innovative Research, Tokyo Institute of Technology

Tokyo 152-8552, Japan

In the Letter [1], Yang et al. reported on an elegant topological vortex laser and proposed that the near-field spin and OAM of the topological edge mode lasing have a one-to-one far-field radiation correspondence [1]. The near-field information is based on frequency dispersions of the topological edge modes, without supporting measurements and/or computer simulations. Unfortunately, their frequency dispersions shown in Fig. 1(c) (see also Fig. S6 and Eqs. (5.3) and (5.4) in Supplemental Material) are wrong. As the result, the mode assignment of the main mode $|-2,+\rangle$ investigated in the Letter is mistaken, which should be $|2,+\rangle$. This spoils the one-to-one correspondence claimed in the Letter.

Dispersion relations of topological interface states are governed by the intrinsic property of the seminal Jackiw-Rebbi state [2], which in the present case emerges from the mass-sign change in the Dirac dispersion associated with the $p-d$ band inversion in the honeycomb-type photonic crystals (PhC) [3]. For the case of the Letter, where a trivial PhC is put at the center and cladded by a topological one, the dispersion relations of the cavity modes formed by the helical interface states are given as follows [4,5]:

$$|l,+\rangle = e^{il\phi}(|p_+\rangle + e^{-i\phi}|d_+\rangle)F(r); \quad \omega = \omega_0 - l\delta\omega \qquad (1)$$

$$|l,-\rangle = e^{il\phi}(|p_-\rangle + e^{i\phi}|d_-\rangle)F(r); \quad \omega = \omega_0 + l\delta\omega \qquad (2)$$

where $l$ stands for OAM, $\pm$ in $|l,\pm\rangle$ for the pseudospin-up and -down states respectively, $\phi$ for the azimuthal angle, $F(r)$ for the radial distribution of wavefunction peaked at the interface between the topological and trivial photonic crystals, $\omega_0$ for the center of the photonic bandgap, and $\delta\omega$ for the frequency separation of ring-cavity modes.

Equations. (1) and (2) manifest the correct spin-momentum locking, where the group velocities are opposite to those shown in Fig. 1(c) (see also Fig. S6 and Eqs. (5.3) and (5.4) in Supplemental Material) in the Letter [1]. As seen in dispersion relation (1), mode $|-2,+\rangle$ takes a higher frequency than mode $|2,+\rangle$, indicating clearly that the mode assignment in the Letter is wrong (see Fig. S6 in

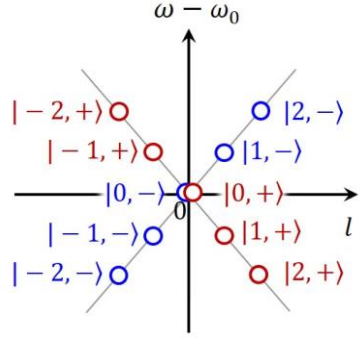

Fig. 1: Dispersion diagram for the ring-cavity modes formed by the helical photonic interface states in a laser device where a trivial PhC at the center is cladded by a topological one. The two modes $|0,\pm\rangle$ are shifted slightly in the horizontal direction for clarity.

Supplemental Material) [1]. The near-field patterns of cavity modes $|2,+\rangle$ and $|-2,+\rangle$ are very different, showing $6\pi$ and $-2\pi$ phase winding around the cavity respectively as can be read from the mode expression in Eq. (1), which can be confirmed by measurements and/or computer simulations.

The dispersion diagram associated with the dispersion relations in Eqs. (1) and (2) is presented Fig. 1. The frequency degeneracy between $|l,+\rangle$ and $|-l,-\rangle$ could be lifted due to the geometric shape of laser device, and next-nearest-neighbor interactions may slightly modify the dispersion relations. Nevertheless, the dispersion diagram in Fig. 1 still provides a useful frame for mode assignment.

We point out that the frequency dispersion diagram shown in Fig. 1 (c) (see also Fig. S6) of the Letter is unphysical, where pseudospin-up and -down modes of the same OAM, such as ($|0,+\rangle$, $|0,-\rangle$) and ($|1,+\rangle$, $|1,-\rangle$) etc., do not take the same horizontal coordinates [1].

Finally, we mention that for a cavity design with topological PhC at the center and cladded by a trivial one, the group velocities should be reversed [4,5]. All the above discussions are valid for honeycomb-type PhC of both dielectric cylinder array buried in air [4,5] and airhole array in dielectric slabs [6].

This work is partially supported by CREST, JST (Core Research for Evolutionary Science and Technology, Japan Science and Technology Agency) under the grant number JPMJCR18T4.

References:
[1] Z.-Q. Yang, Z.-K. Shao, H.-Z. Chen, X.-R. Mao and R.-M. Ma, Phys. Rev. Lett. **125**, 013903 (2020).
[2] R. Jackiw and C. Rebbi, Phys. Rev. D **13**, 3398-3409 (1976).
[3] L.-H. Wu and X. Hu, Phys. Rev. Lett. **114**, 223901 (2015).
[4] G. Siroki, P. A. Huidobro and V. Giannini, Phys. Rev. B **96**, 041408 (2017).

[5] X.-C. Sun and X. Hu, https://arxiv.org/abs/1906.02464 (2019).

[6] S. Barik, H. Miyake, W. DeGottardi, E. Waks and M. Hafezi, New J. Phys. **18**, 113013 (2016).